\documentstyle[12pt]{article}
\global\arraycolsep=2pt

\newcommand{\k}{\vec{k}_{\perp}^2}
\newcommand{\be}{\begin{eqnarray}}
\newcommand{\ee}{\end{eqnarray}}
\newcommand{\la}{\langle}
\newcommand{\ra}{\rangle}

\newcommand{\Dm}{\vec{iD}_{\mu} }
\begin{document}

\begin{titlepage}

\vspace{0.3cm}
\begin{center}
\Large\bf The Pion Form Factor.\\  
  Where Does It Come From ?
 \end{center}

\vspace {0.3cm}

 \begin{center} {\bf Ariel R. Zhitnitsky\footnote{
On leave of absence from Budker Institute of Nuclear Physics,\\ 
Novosibirsk,630090,Russia.\\
e-mail addresses:arz@mail.physics.smu.edu, ariel@sscvx1.ssc.gov}}
 \end{center}
 
\begin{center}
{\it Physics Department, SMU , Dallas, Texas, 75275-0175}

\end{center}
\begin{abstract}

We discuss the nonleading, ``soft"
contribution to the pion form factor at intermediate momentum
transfer within operator product
expansion approach.  We argue, that the corresponding contribution
can  temporarily {\bf simulate} the leading twist behavior
in the extent region of $ Q^2:~~3 GeV^2\leq Q^2\leq 35 GeV^2 $, where
$Q^2 F(Q^2)\sim const.$  
 Such a mechanism, if it is correct, would be an explanation of the
phenomenological success of the dimensional
counting rules (which theoretically correspond to the  
keeping of the asymptotically leading terms only) 
 at available, very modest $Q^2$. The relation  with
quark model calculation is also discussed.

\vskip 0.3cm
 \end{abstract}
\end{titlepage} 
 
{\bf 1. Introduction}
 \vspace {0.3cm} 

The investigation of the exclusive asymptotic processes has
 a long history.
 In early seventies the famous dimensional
counting rules were proposed \cite{Matveev}. 
 The predictions of these rules
agree well with experimental data, such as the
 pion and nucleon form factors,
large angle elastic scattering
cross sections and so on. This agreement served as a stimulus for
further theoretical investigations. The modern approach
to exclusive processes was started
  in the late seventies
and  early eighties  \cite{Brod}. We refer to the review papers 
\cite{Cher},\cite{Brod1} 
  for the  detail   analysis and discussions.

However, since 1981, the applicability of the approach 
\cite{Brod}- \cite{Brod1} at experimentally 
accessible momentum transfers was questioned \cite{Ditt},\cite{Isgur}.
In these papers  
 it was demonstrated, that  the perturbative, asymptotically leading
contribution, is much smaller than the nonleading contributions.
Similar conclusion, supporting this result, came from
the different side, from  the QCD sum rules,
\cite{Rad},\cite{Smilga}, where the direct
calculation of the form factor has been presented
 at $Q^2\leq 3GeV^2$.
  This method, unfortunately, by some technical reasons,
 can not directly be applied for the analysis 
of the form factor $F_{\pi}(Q^2)$ (see, however, \cite{Rad1})
at  larger $Q^2\geq 3GeV^2$. Thus, the 
question `` what kind of contributions are responsible
for the $F_{\pi}(Q^2)$ at $Q^2\geq 3GeV^2$ " can not be answered
within this method.

At the same time, the information which can be extracted from the 
different QCD sum rules, unambiguously shows the asymmetric form
of the leading twist $\pi$ -meson wave function $wf$, \cite{Cher2}.
The application of this $wf$ 
to different amplitudes gives very sensible result 
at $Q^2\simeq 10 GeV^2$  
and one could think that the  region
$3 GeV^2\leq Q^2\leq 10 GeV^2$  is the transition zone,
where  the asymptotically 
leading contribution comes into the game.
However, the recent papers \cite{Ster},\cite{Kroll} do not support
this conjecture. Namely, it was shown that the inclusion
of the intrinsic $\k$- dependence and Sudakov suppression
leads to the self-consistent calculation, but the 
obtained magnitude is too
small (at least  there is factor 3 in the 
description of the $\pi$-meson form factor at $Q^2\simeq 10GeV^2$)
 with respect to the data even if $\phi_{CZ}$ asymmetric
function is used.

Now we are ready to formulate the question,
which we want to discuss in the present letter.

$\bullet$  If the asymptotically leading contribution can not provide
the experimentally observable absolute 
values, than {\it how   can one
explain the very good agreement the 
experimental data with dimensional counting rules}
\footnote{ Let us remind, that these rules
unambiguously predict the dependence of amplitudes on dimensional
parameters. In particular, $Q^2 F_{\pi}(Q^2)\simeq constant.,~~
Q^4 F_{p}(Q^2)\simeq constant.,~~
s^7\frac{d\sigma}{dt}(\gamma p\rightarrow\pi^+n)=f(t/s)$
.. The experimental data are in a good agreement with
these predictions 
in the large region of $s, Q^2 $.at very modest energy and momentum 
transfer. },
which supposed to be valid  only in the region where the  
leading terms dominate? 

It is clear, that the possible explanation can not be related to the
specific amplitude, but instead, it should be
connected, somehow, with the nonperturbative 
wave functions  of the light hadrons ($\pi, \rho, p...$)
which enter to the formulae for exclusive processes.
The analysis of the $\pi$ meson
form factor, presented below supports this idea.
 
To anticipate the events we would like  to formulate here the result
of this letter.
The very unusual properties of the transverse momentum distribution
of the nonperturbative $\pi $ meson wave function lead to the
{\bf temporarily simulation} of the dimensional counting rules
by soft  mechanism for the $F_{\pi}(Q^2)$
at the extent range of intermediate momentum transfer:
$3 GeV^2\leq Q^2\leq 35 GeV^2 $. In this region   the soft contribution
 to $F_{\pi}(Q^2)Q^2   $ does not fall-off, as
naively one could expect, and we estimate it as
$F_{\pi}(Q^2)Q^2\simeq 0.4 GeV^2.   $
 The leading twist contribution,
after Sudakov suppression, gives, according to \cite{Kroll},
a little bit less (about $0.2 GeV^2 $).
 
\vspace {0.3cm}
{\bf 2.Constraints on the nonperturbative wave function $\psi(\k, x)$.}
 \vspace {0.3cm}

First of all let me remind some essential definitions and
 results from \cite{Zhit1}
about nonperturbative $wf$.
We define the pion
axial wave function 
in the following gauge-invariant way:
\be
\label{d}
if_{\pi}q_{\mu}\phi_A (zq,z^2)=
\la 0|\bar{d}(z)\gamma_{\mu}\gamma_5 
e^{ig\int_{-z}^z A_{\mu}dz_{\mu}} u(-z)|\pi(q)\ra \\
\nonumber 
=\sum_n \frac{i^n}{n!}\la 0|\bar{d}(0)\gamma_{\mu}\gamma_5
(iz_{\nu}\stackrel{\leftrightarrow}{D_{\nu}})^n u(0)|\pi(q)\ra ,
\ee
where 
$\stackrel{\leftrightarrow}{D_{\nu}}\equiv
\stackrel{\rightarrow}{D_{\nu}}-\stackrel{\leftarrow}{D_{\nu}}$ and 
$\Dm=i\vec{\partial_{\mu}}+gA_{\mu}^a\frac{\lambda^a}{2}$ is the
covariant derivative. 
From its definition is clear that the set of different
$\pi$ meson matrix elements defines the nonperturbative wave
function. The most important (at  asymptoticaly high $Q^2$) is
the part related to longitudinal distribution. 
The corresponding problem was   studied in \cite{Cher2,Cher}. 
At this paper we are mainly interested in transverse distribution
part, which is very important at available, not very high $Q^2$.
We define the  
mean values of the quark transverse distribution  
in the following way:
\be
\label{1}
\la 0|\bar{d}\gamma_{\nu}
\gamma_5 (\Dm t_{\mu})^{2n} u|\pi(q)\ra=if_{\pi}q_{\nu}
 (-t^2)^n\frac{(2n-1)!!}{(2n)!!}\la \vec{k}_{\perp}^{2n} \ra_A,
\ee
where  transverse vector $t_{\mu}=(0,\vec{t},0)$ is perpendicular
 to the hadron momentum $q_{\mu}=(q_0,0_{\perp},q_z)$. 
The factor $\frac{(2n-1)!!}{(2n)!!}$
is related to the integration over $\phi$ angle in the transverse plane:
$\int d\phi (\cos\phi)^{2n}/ \int d\phi= {(2n-1)!!}/{(2n)!!}$.

We interpret the $\la \vec{k}_{\perp}^{2} \ra$ in this equation
as  a mean value of the quark perpendicular momentum. Of course it
is different from the naive, gauge dependent  definition like
$\la 0|\bar{d}\gamma_{\nu}
\gamma_5 \partial_{\perp}^2 u|\pi(q)\ra $,
because the physical transverse gluon is participant 
of this definition.
However, in QCD the only questions which  
allowed to be asked look like that: 
"what the magnitude of this gauge invariant matrix element?".
Our definition (\ref{d}), (\ref{1}) does satisfy to this requirement.
We believe that such definition is the useful generalization
of the transverse momentum conception for the interactive quark
system.

We can not prove that the mean value
of the transverse momentum distribution $\la\k\ra$,
defined above in terms of QCD 
matrix elements  coincides with the perpendicular
momentum in simple quark model, which   is frequently
used in the phenomenological analysis\footnote{
In order to prove such identification we need to 
demonstrate the confinement
and chiral symmetry breaking , to calculate               
 the meson masses   and wave functions 
and finally, to derive an effective low-energy quark model
in terms of the fundamental QCD parameters (like chiral condensate).
We do not have such ambitious purposes at the moment. Instead,
we assume that such kind of correspondence with intuitive
quark model, takes place with   high enough accuracy.}.
However we would like to make such assumption in 
the following section devoted
to  the  phenomenological analysis of the pion
form factor. At the same time, in this section                                                          
 we prefer to   stick around QCD with its lanquage of operators,
matrix elements and vacuum condensates in order to make 
difference between pure QCD results (without any additional assumptions)
and phenomenological analysis ( next section), where such kind of
assumption, 
  motivated by quark model, has been made.                                                                                    

 It turns out, that the calculation of $\la \vec{k}_{\perp}^{2n} \ra$,
defined in (\ref{1}),
can be reduced to the problem of calculating of the mixed 
vacuum condensates.
In particular
\be
\label{2}
\la \vec{k}_{\perp}^{2} \ra=\frac{5}{36}\frac{\la 
\bar{q}ig\sigma_{\mu\nu} 
 G_{\mu\nu}^a\frac{\lambda^a}{2} q \ra }{\la \bar{q}q\ra}
\simeq \frac{5 m_0^2}{36}\simeq 0.1GeV^2, ~~~m_0^2\simeq 0.8 GeV^2.
\ee
\be
\label{}
\frac{\la \vec{k}_{\perp}^{4}\ra  }{\la \vec{k}_{\perp}^{2}\ra^2}
\simeq 3 K~~\frac{ \la g^2G_{\mu\nu}^aG_{\mu\nu}^a\ra}{m_0^4} 
\simeq  5\div 7,  \nonumber
\ee
where we have introduced the coefficient of nonfactorizability $K$
($K=1$ if the factorization would work) of the  mixed vacuum
condensates of dimension seven. This coefficient can be estimated with
  high  enough accuracy and we expect $K\simeq 2.5\div 3$.

Few comments are in order.
We {\bf define} the nonperturbative wave function $\psi(\k, \xi)$,
which is Fourier transform of the $\phi_{A}(zq,z^2)$,
through its moments which can be expressed in terms of nonperturbative
vacuum condensates. As is known, the condensates are defined in
such a way, that all
 gluon's and quark's
 virtualities, smaller than some parameter $\mu$ 
(point of normalization)
are hidden in the definition of the ``nonperturbative vacuum matrix
elements". All virtualities larger than that should be taken into
account perturbatively, see \cite{Shif2} for detail discussion of
this problem. At the same time,  there is a
relation between moments and condensates,   as explained above.
Thus, all
transverse moments are defined in the same way as condensates do.
This is, actually, the {\bf definition of the nonperturbative
wave function $ \psi( \k, \xi)$}, through its moments
which can be expressed
in terms of nonperturbative vacuum condensates.
  The important consequence of the definition can be 
formulated in the following way: The nonperturbative $wf$
even for sufficiently large $\k$ does not behave like
$1/\k$ as is frequently assumed.
 
Now we want to formulate some additional constraints on the 
nonperturbative wave function $\psi(\k, x)$ in order to model it.
As is known, the knowledge of a finite number of moments is
not sufficient to completely determine the $\psi(\k, x)$; 
the behavior of the 
asymptotically distant terms is  a very important thing as well.
 We assume, that the $\pi$-meson fills
a finite duality interval in the   dispersion
 relation at any $n$,
where $n$ is the order of moment: 
$\la \vec{k}_{\perp}^{2n} \ra $. It gives the following constraint:
  \be
\bullet  ~~~~~~~~~~~~~~~~~~~~~~~ 
\int d\k 
\vec{k}_{\perp}^{2n}\psi(\k, \xi\rightarrow\pm 1)\sim (1-\xi^2)^{n+1}
,~~~ x=\frac{1+\xi}{2} ~~~.\nonumber
   \ee
For the $n=0$ we reproduce the well-known result \cite{Cher2}
for the $\phi$ wave function: $\phi (\xi\rightarrow\pm 1)=
\int d\k \psi(\k, \xi\rightarrow\pm 1)\sim (1-\xi^2) $.
This constraint is extremely important and implies that the $\k$
dependence of the   $\psi(\k, \xi=2x-1)$ comes 
{\bf exclusively in the combination}
$ \k/(1-\xi^2)$ at $\xi\rightarrow\pm 1$.  The byproduct of this 
constraint can be formulated as follows. The standard assumption
on factorizability of the $\psi(\k, \xi ) =\psi(\k )\phi(\xi)$
{\bf does contradict} to the very general property of the theory
formulated above.

Our last constraint comes from the analysis of the 
asymptotically distant
terms. It was argued in \cite{Zhit1} that the moments
$\la \vec{k}_{\perp}^{2n} \ra $
at     asymptotically large $n$  behave in the following
way:
 \be 
\bullet  ~~~~~~~~~~~~~~~~~~~~~~~~~~~
 \int d \xi\int d \k   \psi(\frac{\k}{1-\xi^2},\xi)
( \vec{k}_{\perp}^{2} )^n 
\Rightarrow 1,~~~~~~~~~~~~~~~~~~~` \nonumber
\ee 
which means that the $\psi(\frac{\k}{1-\xi^2})\Rightarrow
\delta(\frac{\k}{1-\xi^2}-S)$ for sufficiently large $\k$
\footnote{From the physical point of view it is clear that such
$\delta$ dependence should be somehow regularized. It turns out that
the large $\alpha_s^n$ corrections are responsible for
this regularization (\cite{Zhit4}, be published ). 
The problem of regularization is not an 
essential point at the moment.};
the    $S$ is some input dimensional parameter,
which will be expressed in terms of $\la\k\ra$.

We want to model the simplest version of the wave function which meets all 
requirements formulated above. First of all, as was explained,
the $\k$ dependence comes only in the combination
$\psi(\frac{\k}{1-\xi^2})$  and  at
sufficiently large $\k$ we have to have  the 
 $\delta(\frac{\k}{1-\xi^2}-S)$ 
-like dependence. 
 Besides that, in order to reproduce
the noticeable fluctuations of
the $\k$ (\ref{2}),
 we have to spread out
the distribution function
between $\k\sim 0$ and $\k\sim S$ in such a way, that the overall area
will be the same, but moments should satisfy to this requirement. 
  In principle, it can be done in arbitrary way. The simplest way
to make the $wf$ wider is to put another $\delta$ function
at $\k=0$. 
 With these remarks in mind we propose the following "two-hump" 
 nonperturbative wave function which meets all requirements 
discussed above:
\be
\label{3}
\psi(\k,\xi)=[A\delta(\frac{\k}{1-\xi^2}-S)+B\delta(\frac{\k}{1-\xi^2})]
[g(\xi^2-\frac{1}{5})+\frac{1}{5}]   \nonumber
\ee
\be
\la\vec{k}_{\perp}^{4}\ra
 \simeq 5\la\k\ra^2 \Rightarrow A=7/8, ~~A+B=\frac{15}{4}~~ g=1,~~
 S=\frac{15}{2}\la\k\ra\simeq 0.8 GeV^2
\nonumber
\ee
Few comments are in order. First,
this is not supposed to be a realistic model.
But we expect that
 this function illustrates features which $\psi(\k,\xi)$
probably does exhibit.
 We put the common factor
 $[g(\xi^2-\frac{1}{5})+\frac{1}{5}] $ in the 
front of the formula  in order to reproduce  the light 
cone   $\phi(\xi)$ function with arbitrary $\la\xi^2\ra$.
For $g=0$ it corresponds to the asymptotic $wf : \phi(\xi)
=3/4 (1-\xi^2)$. For   $g=1$ we reproduce the $\phi(\xi)_{CZ}=
15/4 (1-\xi^2)\xi^2$ with $\la \xi^2\ra\simeq 0.43$.
 
 Before to consider some applications, let me stress, that
  we discuss in this letter some qualitative
properties of the $wf$.   Thus, the model $wf$ proposed above  should
be considered as illustrative example only. We try to    emphasize on
the very unusual qualitative feature of the model. We expect, 
however, that this feature will survive
in the full picture, after physical regularization will be made.

 \vspace {0.3cm}
{\bf 3. The Pion Form factor.}
 \vspace {0.3cm}

The starting point is the famous Drell-Yan formula 
\cite{DY} (for modern, QCD- motivated  employing of this formula,
see \cite{Brod1}),
where the $F_{\pi}(\vec{q}_{\perp}^2)$ is expressed in terms of full
wave functions:
\be
\label{4}
F_{\pi}(\vec{q}_{\perp}^2)=\int\frac{dx_1d^2\vec{k}_{\perp}}
{16\pi^3}\Psi^*_{BL}
(x_1,\vec{k}_{\perp}+x_2\vec{q}_{\perp})
\Psi_{BL}(x_1,\vec{k}_{\perp}),
\ee
where $q^2=-\vec{q}_{\perp}^2$ is the momentum transfer. 
In this formula, the $\Psi_{BL}(x_1,\vec{k}_{\perp})$ is
 the full wave function;
the perturbative tail of $\Psi_{BL}(x_1,\vec{k}_{\perp})$ behaves as
$\alpha_s/ \k$ for large $\k$ and should be taken into account
explicitly in the calculations. This gives the  
 one-gluon-exchange  
(asymptotically leading) formula for the form factor in
terms of soft   pion $wf$ with removed perturbative tail \cite{Brod}.

Let us remind, that   the formula (\ref{4})  takes into account
only the valence Fock states. Besides that, $\k$ in this formula
is the usual (not covariant) 
 perpendicular momentum of the constituents, and not the
mean value $\la\k\ra$   defined in QCD, as a gauge invariant object.
However we make the 
{\bf assumption} that   it is one and the same
variable. The physics behind of it
  can be explained in the following way.

In the formula (\ref{4})   we effectively take into account 
 some gluons (not all of them), which inevitably
 are participants of our definition of $wf$. These gluons mainly carry the 
transverse momentum (which anyhow, does not exceed QCD scale
of order $\mu\sim 1 GeV$) or small ammount of longitudinal momentum.
The contributions of the 
gluons carrying the {\it finite}
longitudinal momentum fraction  are neglected in (\ref{4}).
This is the main assumption.
 It can be justified 
by the direct calculation  \cite{Cher}
   of quark-antiquark-gluon 
(with
 finite momentum fraction)
contribution to $\pi$ meson form factor at large $Q^2$ within the 
standard technique of the operator product expansion. 
By technical reasons
the corresponding calculation has not been  completed, however 
it was found
that the characteristic scale which enters into the game is
of order $1 GeV^2$. Thus, it is very unlikely
to  expect that these contributions
might be important at $Q^2\sim 10 GeV^2$. The second calculation,
which confirms this point, comes from the light cone QCD sum
rules \cite{Braun}. This is almost model independent calculation
demonstrates that the quark-antiquark-gluon
(with
 finite momentum fraction)
contribution does
not exceed $ 20\% $ at available $Q^2$. 

Thus, we expect, that by taking into account
the only "soft'' gluon contribution  (hidden
in the definition of $\k$ (\ref{1})), we catch  the main effect.
Again, there is no proof for that within QCD,
and the only argumentation which can be delivered now in favor of it,
is based on the intuitive picture of quark model, where 
current quark and  soft gluons form a constituent quark 
with nonzero mass and with original quantum numbers.
No evidence where   a gluon would play the role of a valent participant   
with a finite ammount of momentum, is found. 

From the viewpoint of the operator product expansion, the 
assumption formulated above, corresponds to {\bf summing up} a
subset of higher-dimension power corrections. This subset
actually is formed from the infinite number of soft gluons
and  unambiguously
singled out by the definition of nonperturbative $wf$ (\ref{1}).
  
In the following, we preserve the notation 
$\Psi_{BL}(x_1,\vec{k}_{\perp})$ for 
the nonperturbative, soft part  only. It should not confuse
the reader.

The formula (\ref{4}) is written
in terms of Brodsky and Lepage notations \cite{Brod1}; 
 the relation  to our wave function $\psi(\xi,\vec{k}_{\perp})$
looks as follows:
\be
\label{5}
\Psi_{BL}(x_1,\vec{k}_{\perp})=\frac{f_{\pi}16\pi^2
}{\sqrt{6}}\psi(\xi,\vec{k}_{\perp}),~~
\int\frac{ d^2\vec{k}_{\perp}}{16\pi^3} 
 \Psi_{BL}(x_1,\vec{k}_{\perp}) 
=\frac{f_{\pi}}{\sqrt{6}}\phi(\xi), 
\ee
where $f_{\pi}=133 MeV$.
There are two, physically different contributions to the form factor.
We want to separate them and discuss independently.

The first contribution is the standard one, in a sense that it is 
determined by the $\k$- distribution
about $\k=0$ region.  It is proportional
to the   $ B^2$ (we call the corresponding contribution $F_{\pi}^{BB}$)
 and  actually depends very
   strongly on regularization 
procedure. However, at $\vec{q}_{\perp}^2 \simeq few GeV^2$
this contribution   begins to fall off very rapidly because
of the proportionality to $c\la\k\ra^2/\vec{q}_{\perp}^4$
 with some coefficient $c $,
determined by characteristic scale of  the $wf$.  
 Besides that, we expect
that all Fock states are
 equally important in this region   
and  thus, any predictions are model dependent.

However, to give some  qualitative insight on the 
behavior of the contribution of this kind, we estimate
it in the following way.
We regularize our $\delta$ functions in the simplest way-
 we shift its argument:
 $\delta(\frac{\k}{1-\xi^2})\rightarrow 
\delta(\frac{\k}{1-\xi^2}-m^2)$,
by  introducing the  new  effective phenomenological
parameter  $m^2$. This parameter effectively
describes the $\k$- distribution about zero point
and somehow related to $\la\k\ra$. This regularization is
still not sufficient for the calculating of the
integral (\ref{4}) at $q=0$. But we do not need it, 
because we are not going to apply 
this method for the describing of the form factor 
in the vicinity
of $q=0$. Nevertheless, we expect, that at 
$\vec{q}_{\perp}^2\gg m^2$   
only the integral characteristic does matter. Have 
these few comments in mind,  we
estimate the parameter $m^2$ from the requirement of
coincidence theoretical  formula (\ref{4}) with the 
experimental data at 
$\vec{q}_{\perp}^2\simeq 3 GeV^2:~~F_{\pi}\vec{q}_{\perp}^2\simeq
0.4 GeV^2$. It gives $m^2\simeq (0.22 GeV)^2$
for our set of parameters (\ref{3}).
 We display the result of this contribution to the  form factor
$\vec{q}_{\perp}^2F_{\pi}(\vec{q}_{\perp}^2)$ in Fig. 1 by small-dashed curve.
We took into account in this numerical calculation
the evolution of the $wf$ which determined by parameter $g(q^2)$
\cite{Brod}-\cite{Brod1}:
\be
\label{6}
g(q)=g(\mu)(\frac{\alpha_s(q)}{\alpha_s(\mu)})^{50/81},~~~~
g(\mu)=1,~~ \mu =0.5 GeV,~~ \Lambda_{QCD}=100 MeV.
\ee
 Let me stress again: we are not pretending to have made
a reliable calculation of  the form factor here;
we displayed this contribution only for the 
{\bf illustrative}
purposes. The main feature of this contribution-- 
it gives very reasonable
magnitude for the intermediate region about few $GeV^2$ 
and it
starts to fall off very quickly at $\vec{q}_{\perp}^2\geq few ~\la\k\ra$. 
We expect that {\bf any} reasonable, well localized $wf$
with the scale $\sim \la\k\ra $ leads to the same
behavior. We will not discuss this standard  
contribution any more.

 Currently, much more interesting for us, is the   
  interference term, proportional to $AB$.
It is clear, that with increasing of $\vec{q}_{\perp}^2$, the contribution
coming from the overlap of the $\delta$ functions starts to grow.
Because the $\delta$ functions are well separated from each other 
on the value of order $S\simeq 0.8 GeV^2$, we expect
that this contribution will start to increase at 
high enough $\vec{q}_{\perp}^2\gg 4S\simeq 4GeV^2$.
This contribution almost model-independent in a sense that it
does not depend on the infrared regularization parameter
$m^2$ which we introduced before, provided that $m^2\ll S$. Let me remind 
at this point, that our parameters $S$ and $\la\k\ra$ are not the
new phenomenological input parameters, but have been derived 
(\ref{2}, \ref{3}) in terms of
the   vacuum characteristics,    which are  known independently.

With these remarks in mind, we can explicitly calculate this
interference contribution:
\be
\label{7}
F_{\pi}^{AB}(\vec{q}_{\perp}^2)=AB\frac{2\pi^2f_{\pi}^2}{3S}
(1-\xi_0^2)^2[g(\vec{q}_{\perp}^2)\xi_0^2+
\frac{1}{5}(1-g(\vec{q}_{\perp}^2))]^2,
~\xi_0\equiv\frac{\vec{q}_{\perp}^2-4S}{\vec{q}_{\perp}^2+4S}.
\ee
We display the corresponding contribution to the 
$\vec{q}_{\perp}^2F_{\pi}(\vec{q}_{\perp}^2)$ in Fig. 1 by 
the large-dashed curve.
The solid line on this picture corresponds to the total result.

Few comments are in order. As  was expected, the interference term
starts to {\bf grow} at high enough momentum transfer. This is the
{\bf main qualitative effect}.  
The  contribution
under consideration is subject to Sudakov corrections.
An estimate of these corrections reveals that they are small enough. The reason for that is the large scale $4S  $ which enter to the 
formula (\ref{7}); thus the $\xi_0$ is far away from the end point region
$ \pm 1$ even at large $ \vec{q}_{\perp}^2$. 
As the second remark, we note, that the form factor is the 
integral characteristic and so, we expect that any reasonable
regularization of $\delta$ functions which respect constraints,
will not change the $F_{\pi}^{AB}$ drastically.

At the same time we want to note that the result strongly depends
on $\la\xi^2\ra$. We display on Fig.2 three different curves
for $\vec{q}_{\perp}^2F_{\pi}(\vec{q}_{\perp}^2)$
which correspond to different choice of $\la\xi^2\ra$.
The small-dashed curve corresponds to the asymptotic $wf$ with
$\la\xi^2\ra =0.2$ and $g(\mu)=0$. The large-dashed curve 
corresponds to $\la\xi^2\ra =0.3,~g(\mu)=0.44$.
The solid line corresponds to $\la\xi^2\ra =0.4,~g(\mu)=0.88$.
Let us note, that in these calculations we did not change
the infrared regulator, $m^2$ which is an effective parameter
for calculation of the ``standard" $F_{\pi}^{BB}$
contribution.  But we made all necessary substitutions related 
to changing of the parameter $g(\mu)$ in eq.(\ref{3}).
We observe the   strong dependence on parameter
$\la\xi^2\ra$. In particular, at $\la\xi^2\ra=0.4$, the
soft contribution to the form factor  at $\vec{q}_{\perp}^2
=10 GeV^2$ is equal to
 $\vec{q}_{\perp}^2F_{\pi}(\vec{q}_{\perp}^2)\simeq 0.4$.
For the asymptotic $wf$ with $\la\xi^2\ra=0.2$ we have
$\vec{q}_{\perp}^2F_{\pi}(\vec{q}_{\perp}^2)\simeq 0.1$.
These astimations are in a very good agreement with
absolutely independent recent calculations \cite{Braun}, who used
a quite different method.

\vspace {0.3cm}
 {\bf 4. Summary and Outlook.}
 \vspace {0.3cm}

 We hope we proposed the new ``soft" mechanism which could be responsible 
for the  explanation of the phenomenological success of the dimensional
counting rules. The main idea is based on the very unusual
feature 
(noticeable fluctuations
of the transverse momentum ) of the wave function  . It leads to the
``two-hump" shape for the nonperturbative $wf$ in the transverse
direction
and to the appearing of the {\bf new} scale ($S=0.8 GeV^2$) in the
 problem, in addition to the standard low energy parameter
$\la\k\ra\simeq 0.1 GeV^2$
\footnote{ In the standard
quantum mechanics it is very unlikely 
 to have the   
two-particle  lowest state
wave function with  
 such unusual form. 
However, we consider our $wf$ as an effective one, which counts
infinite number of soft gluons. The information about them is hidden
into the definition of the effective transverse momentum.
We believe that precisely these gluons are responsible for
the making constituent quark from the current massless quarks
and gluons.}.
 As the consequence of it, the increasing
of the $\vec{q}_{\perp}^2$ 
in the pion form factor leads to the increasing of overlap
of these humps. It gives the growing contribution
at intermediate $\vec{q}_{\perp}^2$
(  parametrically it  falls off very quickly, of course).
Together with the standard, decreasing contribution, it 
could {\bf simulate} the leading twist behavior in the
extent region of $\vec{q}_{\perp}^2$. Our conclusions
 essentially support the picture described in \cite{Rady},
but from the quite different side.

It would 
be interesting to check the conjecture formulated above
(about an equivalence of the  transverse momentum
defined in terms of the gauge invariant operatorors in 
QCD and perpendicular momentum in quark model)
in the  phenomenological analysis for the  different amplitudes.
 Besides that,
we  would expect that the same feature might occur in the nucleon
$wf$ and we hope to discuss it somewhere else. 
I thank Tolya Radyushkin and Stan Brodsky
 for useful comments and insights
on the number of subjects related to this letter.

 This work is supported by the Texas National Research 
Laboratory Commission under  grant \# RCFY 93-229.

\newpage
{\bf FIGURES}

\vspace{0.3cm}

Fig.1
Small-dashed line is the $F^{BB}$ contribution to the pion form factor
$\vec{q}_{\perp}^2 F_{\pi}^{BB}(\vec{q}_{\perp}^2)$. 
Large-dashed line is the interference term, 
$\vec{q}_{\perp}^2 F_{\pi}^{AB}(\vec{q}_{\perp}^2)$.
Solid line is the total result 
$\vec{q}_{\perp}^2 F_{\pi}(\vec{q}_{\perp}^2)$
for $g (\mu)=1$.

\vspace {0.3cm}

Fig.2
Small-dashed line is the 
$\vec{q}_{\perp}^2 F_{\pi}(\vec{q}_{\perp}^2)$
at $\la\xi^2\ra=0.2$; Large-dashed line corresponds to
$\la\xi^2\ra=0.3$; Solid line is the result for
$\la\xi^2\ra=0.4$.
  
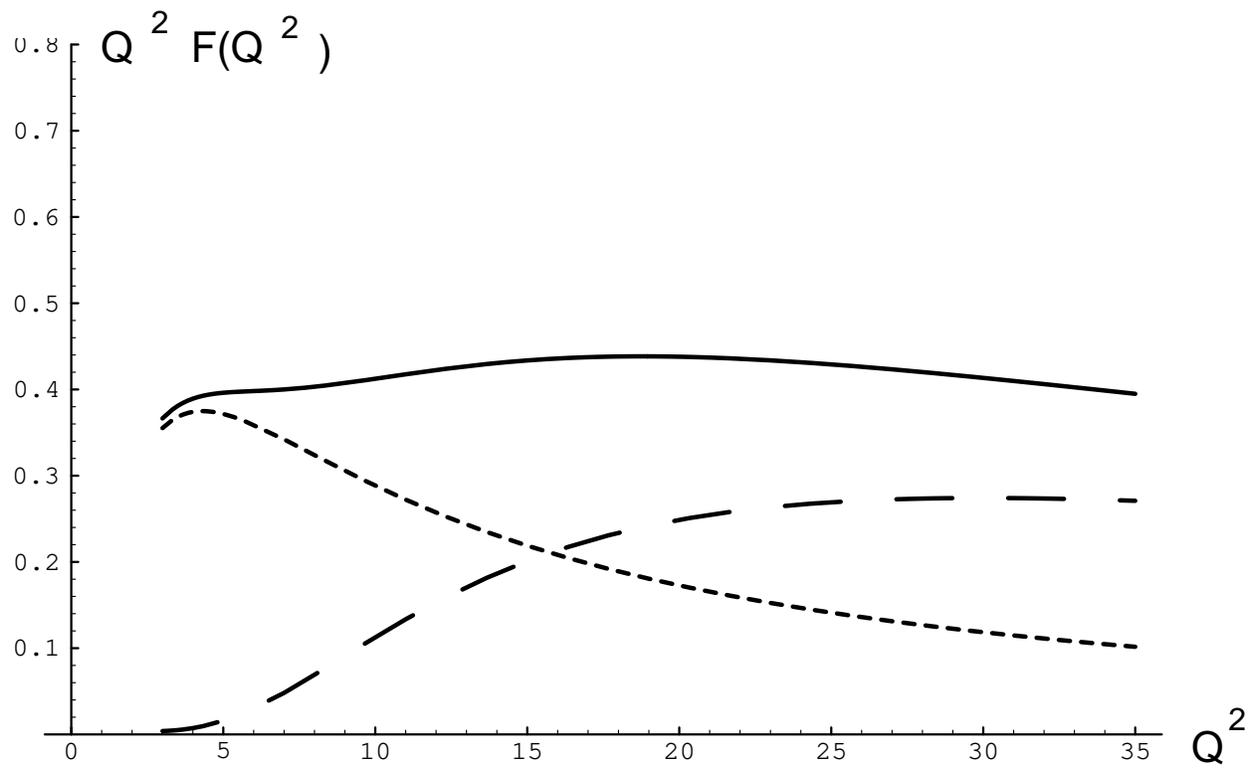
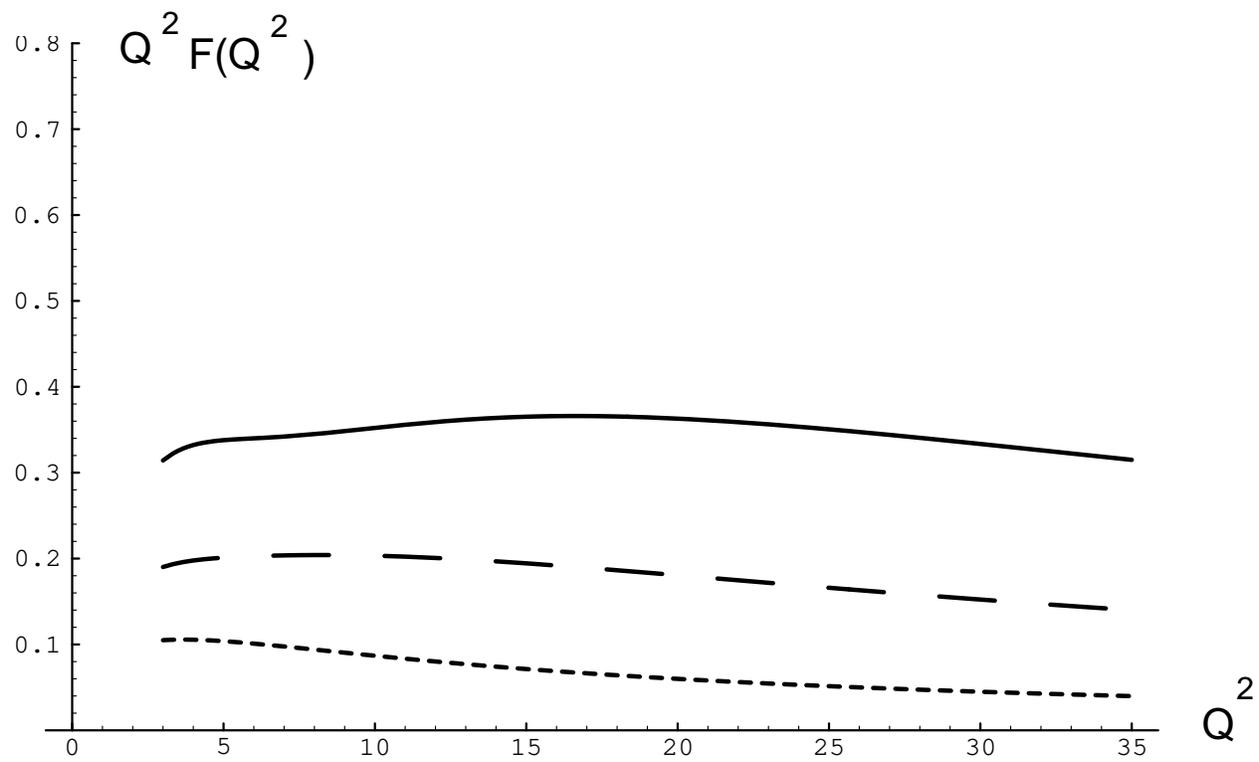
\end{document}